\documentclass[runningheads]{llncs}

\usepackage[T1]{fontenc}
\usepackage{times}

\usepackage{amsmath}
\usepackage{amssymb}  % 推荐
\usepackage{amsfonts}
\usepackage[breaklinks,colorlinks,allcolors=blue]{hyperref}
\usepackage[font=small,skip=2pt]{caption}
\usepackage{booktabs} % for \toprule, \midrule, \bottomrule
\usepackage[normalem]{ulem} % for \uline (better underlines)
\usepackage{array} % for column spacing control
\usepackage{lipsum}
\usepackage{tikz}
\usepackage{graphicx}
\usepackage{siunitx}
\usepackage{subcaption}
\usepackage{threeparttable}
\usepackage{adjustbox}
\usepackage{etoolbox}
\usepackage{float}
\usepackage{placeins}
\usepackage{bbding}
\usepackage[font=small,labelfont=bf]{caption}

\makeatletter
\renewcommand\section{\@startsection{section}{1}{\z@}%
  {-1ex plus -0.5ex minus -0.3ex}%
  {0.5ex plus 0.3ex}%
  {\normalfont\large\bfseries}}
\renewcommand\subsection{\@startsection{subsection}{2}{\z@}%
  {-0.8ex plus -0.3ex minus -0.3ex}%
  {0.4ex plus 0.3ex}%
  {\normalfont\normalsize\bfseries}}
\makeatother

\newcommand{\tablesetup}{%
  \small
  \setlength{\tabcolsep}{5pt}% column padding
  \renewcommand{\arraystretch}{1.15}% row height
}
\usetikzlibrary{positioning}
\newcommand{\TriptychRow}[1]{%
\includegraphics[width=0.99\linewidth,trim= 50 60 50 50,clip]{#1}
}

\title{Joint Neural SDF Reconstruction and Semantic Segmentation for CAD Models}
\author{Shen Fan\inst{1}\Envelope \and Przemyslaw Musialski\inst{1}}
\authorrunning{S. Fan and P. Musialski}
\institute{New Jersey Institute of Technology, Newark NJ 07102, USA \\
\email{\{sf269,przem\}@njit.edu}} % \and Anonymous University \and Anonymous Organization\\
\begin{document}
\maketitle

\begin{abstract}
We propose a simple, data-efficient pipeline that augments an implicit reconstruction network based on neural SDF-based CAD parts with a part-segmentation head trained under \emph{PartField-generated} supervision. 
Unlike methods tied to fixed taxonomies, our model accepts \emph{meshes with any number of parts} and produces coherent, geometry-aligned labels in a single pass. 
We evaluate on randomly sampled CAD meshes from the ABC dataset with intentionally varied part car\-di\-na\-li\-ties---including over-segmented shapes---and report strong performance across reconstruction (CD$_{L1}$/CD$_{L2}$, F1$_\mu$, NC) and segmentation (mIoU, Accuracy), together with a new \emph{Segmentation Consistency} metric that captures local label smoothness. We attach a lightweight segmentation head to the FlatCAD SDF trunk; on a paired evaluation it does not alter reconstruction while providing accurate part labels for meshes with any number of parts.
Even under degraded reconstructions on thin or intricate geometries, segmentation remains accurate and label-coherent, often preserving the correct part count. 
Our approach therefore offers a practical route to semantically structured CAD meshes without requiring curated taxonomies or exact palette matches. We discuss limitations in boundary precision---partly due to per-face supervision---and outline paths toward boundary-aware training and higher resolution labels.

\end{abstract}

\section{Introduction}

Neural implicit fields, particularly signed distance functions (SDFs), are widely used to represent 3D geometry by encoding a surface as the zero-level set of a continuous neural function~\cite{Park2019DeepSDFLC,Mescheder2018OccupancyNL,Sitzmann2020ImplicitNR,Mller2022InstantNG}. This formulation provides resolution-independent evaluation and differentiable access to surface normals, and it has been adopted for reconstruction from point clouds and scans.

For computer-aided design (CAD) data, additional structure is required. CAD parts are predominantly composed of simple primitives (e.g., planes, cylinders, cones) and blends joined by sharp feature curves. Training with point-wise losses alone can yield surfaces that interpolate samples yet deviate from these structural regularities. Curvature-aware priors have therefore been introduced: the Implicit Geometric Regularizer (IGR) encourages unit-norm gradients~\cite{Gropp2020ImplicitGR}; DiGS penalizes divergence of the gradient field~\cite{BenShabat2021DiGSD}; NeurCADRecon penalizes Gaussian curvature on a near-surface shell~\cite{Dong2024NeurCADReconNR}; and Neural-Singular-Hessian constrains the Hessian near the surface~\cite{Wang2023NeuralSingularHessianIN}. More recent formulations reduce the cost of curvature regularization by targeting specific mixed second-order terms while preserving geometric behavior~\cite{Yin2025FlatCAD}, or specific activations for feature preservation~\cite{Fan2024_3DRecon}. 

A limitation of these approaches is that they focus on geometry without providing semantic part structure, which is often required in CAD workflows (e.g., for editing or downstream parametric recovery). Mesh-based pipelines can provide part labels, but they do not unify semantics with the implicit representation.

This work presents a joint implicit framework for CAD surface reconstruction and part segmentation. A shared neural backbone is paired with two output heads: one predicts SDF values trained with standard data, Eikonal, and curvature-regularization terms; the other predicts per-point semantic logits trained with a segmentation loss. This design allows querying both geometry and part labels in the same continuous 3D field. We also compare alternative segmentation-head designs (e.g., sine-activated and ReLU-based variants) and report their practical trade-offs.

\paragraph{Contributions.} (i) A unified implicit formulation that combines curvature-regularized SDF reconstruction with per-part semantic labeling, and (ii) an empirical analysis of segmentation-head designs within this formulation, including the corresponding training objective and evaluation protocol on CAD models with part-level annotations.

\section{Related Work}

\paragraph{Implicit Neural Representations for 3D Shapes}

Neural implicit representations have revolutionized 3D shape modeling by encoding geometry as continuous functions. Deep\-SDF~\cite{Park2019DeepSDFLC} pioneered learning signed distance functions with neural networks, while Occupancy Networks~\cite{Mescheder2018OccupancyNL} used occupancy fields for shape representation. These foundational works demonstrated that MLPs could store complex geometries in a compact, differentiable form. 
Subsequent research has focused on improving reconstruction quality and efficiency.
% Neural-Pull~\cite{Ma2020NeuralPullLS} optimizes for piecewise smooth surfaces by pulling query points onto the surface. 
DiGS~\cite{BenShabat2021DiGSD} introduces a divergence-guided shape representation for improved accuracy. GIFS~\cite{Ye2022GIFSNI} proposes a general implicit function for arbitrary shape topology. Points2Surf~\cite{Erler2020Points2SurfLI} learns implicit surfaces directly from raw point clouds without normals. 
For improved representation capacity, several works explore hierarchical and local approaches. Deep Local Shapes~\cite{Chabra2020DeepLS} learns a collection of local SDFs for detailed reconstruction. Convolutional Occupancy Networks~\cite{Peng2020ConvolutionalON} incorporates convolutional operations into implicit representations. IF-Net~\cite{Chibane2020ImplicitFI} uses multi-scale deep features for shape completion. 
These methods trade memory for improved detail capture.

\paragraph{CAD-Specific Neural Implicit Methods}

CAD models present unique challenges due to their precise geometric features, sharp edges, and planar surfaces. Standard implicit methods often struggle with these characteristics, motivating specialized approaches. 

Neural-Singular-Hessian (NSH)~\cite{Wang2023NeuralSingularHessianIN} enforces singular Hessian constraints to preserve  features in point cloud reconstruction. The method exploits the mathematical property that sharp edges correspond to singular points in the Hessian matrix. NeurCADRecon~\cite{Dong2024NeurCADReconNR} explicitly targets zero Gaussian curvature regions common in manufactured objects, using specialized losses to encourage flat surfaces. 
FlatCAD~\cite{Yin2025FlatCAD} achieves the state-of-the-art in CAD reconstruction using an efficient curvature regularizer. FlatCAD's key innovation is a proxy curvature term that encourages flat and cylindrical surfaces without expensive second-order derivative computation. Its SIREN backbone makes it an ideal base for our joint segmentation-reconstruction approach.
Other CAD-focused methods include Point2Cyl~\cite{Uy2021Point2CylRE}, which reverse-engineers 3D objects into extrusion cylinders, and various primitive-based approaches that decompose shapes into simple geometric elements. However, these methods typically require strong geometric priors that limit their generality.

\paragraph{3D Part Segmentation: From Manual to Automated Approaches}

While substantial progress has been made in 2D object part segmentation, the 3D counterpart has received less attention, in part due to the scarcity of annotated 3D datasets. 3D object part segmentation is essential for applications ranging from shape editing and retrieval to robotic manipulation and manufacturing. Traditional approaches relied on geometric cues~\cite{shlafman2002metamorphosis}, primitive fitting~\cite{schnabel2007efficient}, or spectral clustering~\cite{sidi2011unsupervised}, but these methods struggle with complex, non-convex shapes.

Learning-based methods have shown promise but typically require extensive labeled data. PointNet~\cite{Qi2017PointNetDH} and its variants process point clouds directly but need category-specific training. BAE-NET~\cite{chen2019bae} learns branched autoencoders for co-segmentation without part-level supervision but is limited to shapes with consistent structure. The lack of large-scale 3D part annotations remains a fundamental bottleneck.

\paragraph{Automated Part Annotation}

PartField~\cite{Liu2025PARTFIELDL3} represents a significant advance in automated 3D part annotation. Unlike methods that rely on 2D projections or require category-specific training, PartField learns generalizable 3D feature fields that respect geometric boundaries. The method can decompose arbitrary meshes into semantically meaningful parts without manual annotation.

Critically, PartField's ability to handle over-segmentation—producing more parts than strictly necessary—provides valuable robustness testing for downstream methods. By applying PartField to the ABC dataset, we create a scalable pipeline for generating part-level supervision without human intervention.

\paragraph{Joint Learning of Geometry and Semantics}

Several works explore joint learning of 3D geometry and semantics, though primarily for scene understanding rather than part segmentation. SurroundSDF~\cite{Liu2024SurroundSDFI3} predicts both signed distance and semantic fields from surround-view images. ObjectSDF~\cite{objectSDF2022} and ObjectSDF++~\cite{objectSDFpp2023} define per-object semantic fields for compositional scene representation. ClusteringSDF~\cite{Wu2024ClusteringSDFSN} fuses noisy 2D semantic predictions into consistent 3D representations.

For part-level understanding, PartSDF~\cite{Talabot2025PartSDFPI} outputs separate SDFs for each part but requires part annotations during training. Our work differs by learning a unified representation with a shared feature extractor, enabling more efficient inference and better feature sharing between geometry and semantics.
% write the method (secondary priority)
\section{Method}

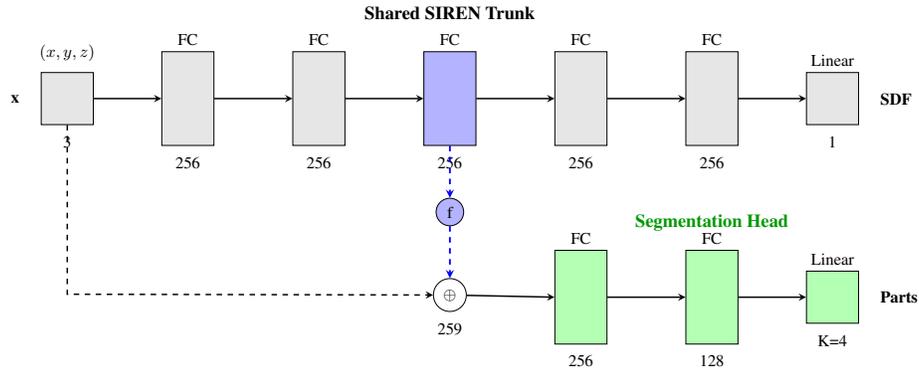
\begin{figure}[t]
\centering
\resizebox{\linewidth}{!}{%
\begin{tikzpicture}[
    scale=1,
    node distance=1.5cm,
    box/.style={rectangle, draw, minimum height=1.8cm, minimum width=1cm, fill=gray!20},
    smallbox/.style={rectangle, draw, minimum height=1cm, minimum width=1cm, fill=gray!20},
    label/.style={above, font=\small},
    arrow/.style={->, thick, >=stealth}
]

% Input
\node[smallbox] (input) at (0,0) {};
\node[above=0.1cm of input, font=\small] {$(x,y,z)$};
\node[below=0.1cm of input, font=\small] {3};
\node[left=0.3cm of input, font=\small] {\textbf{x}};

% Shared SIREN trunk layers
\node[box, right=1.3cm of input] (fc1) {};
\node[label] at (fc1.north) {FC};
\node[below=0.1cm of fc1, font=\small] {256};

\node[box, right=of fc1] (fc2) {};
\node[label] at (fc2.north) {FC};
\node[below=0.1cm of fc2, font=\small] {256};

\node[box, right=of fc2, fill=blue!30] (fc3) {};
\node[label] at (fc3.north) {FC};
\node[below=0.1cm of fc3, font=\small] {256};

\node[box, right=of fc3] (fc4) {};
\node[label] at (fc4.north) {FC};
\node[below=0.1cm of fc4, font=\small] {256};

\node[box, right=of fc4] (fc5) {};
\node[label] at (fc5.north) {FC};
\node[below=0.1cm of fc5, font=\small] {256};

% SDF output
\node[smallbox, right=1.3cm of fc5] (sdf) {};
\node[label] at (sdf.north) {Linear};
\node[below=0.1cm of sdf, font=\small] {1};

\node[right=0.3cm of sdf, font=\small] {\textbf{SDF}};

% Feature extraction point
\node[circle, draw, fill=blue!30, below=1cm of fc3, minimum size=0.5cm] (feat) {\small f};

% Concatenation node - positioned to align with segmentation head
\node[circle, draw, below=1cm of feat, minimum size=0.5cm] (concat) {$\oplus$};
\node[below=0.1cm of concat, font=\small] {259};

% Segmentation head title
\node[below=1.2cm of fc5, font=\normalsize\bfseries, green!60!black] {Segmentation Head};

% Segmentation head boxes - positioned below and aligned with main path
\node[box, below=2cm of fc4, fill=green!30] (seg1) {};
\node[label] at (seg1.north) {FC};
\node[below=0.1cm of seg1, font=\small] {256};

\node[box, right=of seg1, fill=green!30] (seg2) {};
\node[label] at (seg2.north) {FC};
\node[below=0.1cm of seg2, font=\small] {128};

\node[smallbox, right=1.3cm of seg2, fill=green!30] (seg3) {};
\node[label] at (seg3.north) {Linear};
\node[below=0.1cm of seg3, font=\small] {K=4};

\node[right=0.3cm of seg3, font=\small] {\textbf{Parts}};

% Section label for trunk
\node[above=0.5cm of fc3, font=\normalsize\bfseries] {Shared SIREN Trunk};

% Arrows for main SDF path
\draw[arrow] (input) -- (fc1);
\draw[arrow] (fc1) -- (fc2);
\draw[arrow] (fc2) -- (fc3);
\draw[arrow] (fc3) -- (fc4);
\draw[arrow] (fc4) -- (fc5);
\draw[arrow] (fc5) -- (sdf);

% Feature extraction arrow
\draw[arrow, blue, dashed, line width=1pt] (fc3.south) -- (feat);
\draw[arrow, blue, dashed, line width=1pt] (feat) -- (concat);

% Input bypass to concatenation
\draw[arrow, dashed, line width=0.8pt] (input.south) -- ++(0,-3.25) -- (concat.west);

% Segmentation path arrows
\draw[arrow] (concat) -- (seg1.west);
\draw[arrow] (seg1) -- (seg2);
\draw[arrow] (seg2) -- (seg3);

% % Bottom annotation
% \node[below=4.5cm of fc3, font=\small, align=center, text width=14cm] {
%     All FC layers use sine activation\\
%     Segmentation head uses ReLU/SIREN activation with 0.2 dropout
% };

\end{tikzpicture}%
}
\caption{Network architecture of the joint SDF-segmentation network. The shared SIREN trunk (5 layers, 256 neurons each) processes input coordinates through sine activations. Features are extracted from layer 3 (blue) and concatenated with input coordinates before being processed by a separate segmentation MLP (green). The original SDF prediction path remains unchanged from FlatCAD.}
\label{fig:architecture}
\end{figure}

\subsection{Overview and Goals}
We propose an extension of FlatCAD \cite{Yin2025FlatCAD} that jointly learns surface reconstruction and semantic part segmentation from labeled point clouds. While FlatCAD excels at reconstructing flat geometric regions through its off-diagonal Weingarten loss formulation, it lacks the capability for semantic part segmentation. Our method introduces a shared feature learning framework that leverages the geometric representations learned by FlatCAD's SIREN network to simultaneously predict both signed distance values and part labels.

The key insight is that intermediate features learned for accurate surface reconstruction contain rich geometric information that can be effectively repurposed for semantic segmentation. By sharing the feature extraction backbone between tasks while maintaining separate prediction heads, we achieve:
\begin{enumerate}
    \item Accurate surface reconstruction with flat regions (inherited from FlatCAD).
    \item Consistent part segmentation that respects geometric boundaries.
    \item Computational efficiency through shared feature extraction.
    \item Minimal architectural overhead compared to the base FlatCAD model.
\end{enumerate}

\subsection{Network Architecture}
\subsubsection{Shared Feature Network}
As shown in Figure~\ref{fig:architecture}, the backbone is a five-layer SIREN~\cite{Sitzmann2020ImplicitNR}. The input layer maps $\mathbb{R}^3$ to $\mathbb{R}^{256}$ with a sine activation of frequency $\omega_0=30$; four fully connected hidden layers of width $256$ use sine activations with frequency $\omega=1$. Denoting activations by $\phi_i$,
\begin{equation*}
\phi_i(\mathbf{x}) =
\begin{cases}
\sin\!\big(\omega_0(\mathbf{W}_0\mathbf{x}+\mathbf{b}_0)\big), & i=0,\\[2pt]
\sin\!\big(\omega(\mathbf{W}_i\phi_{i-1}+\mathbf{b}_i)\big), & i\in\{1,2,3,4\}.
\end{cases}
\end{equation*}
We tap intermediate features at the third layer, $\mathbf{f}=\phi_2(\mathbf{x})\in\mathbb{R}^{256}$, and reuse them in both heads. $W_i$ and $b_i$ denote learnable weights and biases of the ith fully connected layer.

\subsubsection{Dual-Head Architecture}
\textbf{SDF head.} The reconstruction path continues with two sine layers and a linear projection
\begin{align*}
\phi_3(\mathbf{x}) &= \sin\!\big(\omega(\mathbf{W}_3\phi_2+\mathbf{b}_3)\big),\\
\phi_4(\mathbf{x}) &= \sin\!\big(\omega(\mathbf{W}_4\phi_3+\mathbf{b}_4)\big),\\
f(\mathbf{x}) &= \mathbf{W}_{\text{sdf}}\phi_4+\mathbf{b}_{\text{sdf}},\quad \mathbf{W}_{\text{sdf}}\in\mathbb{R}^{1\times256}.
\end{align*}

\noindent\textbf{Segmentation head.} The segmentation head is a simple three-layer MLP we designed. For part prediction we concatenate $\mathbf{f}$ with the raw coordinates, $\mathbf{f}_{\text{seg}}=[\mathbf{f};\mathbf{x}]\in\mathbb{R}^{259}$, and apply a three-layer MLP: $259\!\rightarrow\!256$ (ReLU, dropout $p=0.2$), $256\!\rightarrow\!128$ (ReLU), and $128\!\rightarrow\!K$ (linear logits; $K=4$ in our experiments). This branch adds $\approx10^5$ parameters (about $25\%$ over a $\sim4{\times}10^5$ backbone).

\subsection{Joint Geometric and Segmentation Loss}

We optimize a joint objective that combines geometric reconstruction terms with a semantic segmentation term. Let
\(\mathcal{X}_{\mathrm{man}}=\{x_i\}_{i=1}^{N}\) denote on-surface (``manifold'') samples with available part labels on a subset
\(\mathcal{X}_{\mathrm{lab}}\subseteq\mathcal{X}_{\mathrm{man}}\),
\(\mathcal{X}_{\mathrm{non}}=\{y_j\}_{j=1}^{M}\) denote uniform off-surface (``non-manifold'') samples,
and \(\Omega=\{p_\ell\}_{\ell=1}^{L}\) denote a thin near-surface shell used for curvature regularization.
The network predicts a signed distance \(f:\mathbb{R}^3\!\to\!\mathbb{R}\) and part logits
\(\mathbf{g}:\mathbb{R}^3\!\to\!\mathbb{R}^K\) (with \(K\) classes).
The total loss is
\begin{equation}
\mathcal{L}_{\mathrm{total}}
=\lambda_{\mathrm{DM}}\mathcal{L}_{\mathrm{DM}}
+\lambda_{\mathrm{DNM}}\mathcal{L}_{\mathrm{DNM}}
+\lambda_{\mathrm{EIK}}\mathcal{L}_{\mathrm{EIK}}
+\lambda_{\mathrm{ODW}}\mathcal{L}_{\mathrm{ODW}}
+\lambda_{\mathrm{SEG}}\mathcal{L}_{\mathrm{SEG}}.
\end{equation}

\paragraph{Manifold (Dirichlet) loss}
We pull on-surface samples to the zero level set~\cite{Atzmon2019SALSA}:
\begin{equation}
\mathcal{L}_{\mathrm{DM}}
=\frac{1}{N}\sum_{x\in\mathcal{X}_{\mathrm{man}}}\bigl|\,f(x)\,\bigr|.
\end{equation}

\paragraph{Non-manifold (sign-agnostic) loss}
For off-surface points we use a rapidly decaying sign-agnostic penalty (with \(\alpha=100\))~\cite{Atzmon2020SALSA}:
\begin{equation}
\mathcal{L}_{\mathrm{DNM}}
=\frac{1}{M}\sum_{y\in\mathcal{X}_{\mathrm{non}}}\exp\!\bigl(-\alpha\,|f(y)|\bigr).
\end{equation}

\paragraph{Eikonal loss.}
We regularize the field to satisfy the unit-gradient property of signed distances~\cite{Gropp2020ImplicitGR}:
\begin{equation}
\mathcal{L}_{\mathrm{EIK}}
=\frac{1}{K_e}\sum_{z\in\mathcal{Z}}
\Bigl(\,\|\nabla f(z)\|_2^2-1\,\Bigr)^2,
\end{equation}
where \(\mathcal{Z}\) is a set of points sampled over the entire domain, and \(K_e=|\mathcal{Z}|\).

\paragraph{Curvature gap loss.}
The off-diagonal Weingarten loss~\cite{Yin2025FlatCAD}, which uses shell samples \(p\in\Omega\), where \(n=\nabla f(p)/\|\nabla f(p)\|\) and \((u,v)\) form an orthonormal tangent frame at \(p\) (any rotation in the tangent plane is admissible) preserves flat regions. 
The off-diagonal entry of the Weingarten map in this frame is
\begin{equation*}
S_{12}(p)=\frac{u^\top H_f(p)\,v}{\|\nabla f(p)\|_2},
\end{equation*}
where \(H_f\) is the Hessian of \(f\).
We penalize its magnitude to suppress the curvature gap:
\begin{equation}
\mathcal{L}_{\mathrm{ODW}}
=\frac{1}{L}\sum_{p\in\Omega}\bigl|\,S_{12}(p)\,\bigr|.
\end{equation}
In practice, \(S_{12}(p)\) is evaluated off-surface on the shell for numerical stability; either a Hessian–vector product or a symmetric finite-difference stencil may be used to obtain \(u^\top H_f(p)\,v\) as proposed in~\cite{Yin2025FlatCAD}. 

\paragraph{Segmentation loss.}
For labeled manifold samples \(x\in\mathcal{X}_{\mathrm{lab}}\) with ground-truth part label \(y(x)\in\{1,\dots,K\}\), we apply cross-entropy (CE) on the logits \(\mathbf{g}(x)\):
\begin{equation}
\mathcal{L}_{\mathrm{SEG}}
=\frac{1}{|\mathcal{X}_{\mathrm{lab}}|}
\sum_{x\in\mathcal{X}_{\mathrm{lab}}}
\mathrm{CE}\!\bigl(\mathrm{softmax}(\mathbf{g}(x)),\,y(x)\bigr).
\end{equation}
No segmentation loss is applied to unlabeled points.

\subsection{Implementation Details}

\paragraph{Inputs and outputs.}
We applied a pretrained PartField model to the ABC dataset~\cite{Koch_2019_CVPR} to produce per-face cluster IDs (visualized as colors) and exported both the colorized mesh and the per-face label array as ground-truth labels for training. From each ground-truth mesh we sample $30{,}000$ surface points with associated part labels. Each iteration draws $20{,}000$ on-surface (manifold) samples, $20{,}000$ uniform off-surface (non-manifold) samples, and a thin near-surface shell used by the curvature regularizer. The network ingests 3D coordinates (and labels when present) and predicts SDF values for all points and $K$-way logits for labeled manifold points.

\paragraph{Curvature regularization and notation.}
We use the FlatCAD curvature-gap loss (off-diagonal Weingarten loss) on the near-surface shell. In the objective its weight is denoted $\lambda_{\text{ODW}}$ (value $10$ in our runs), matching our code and tables.

\paragraph{Initialization.}
Sine layers follow SIREN initialization: input weights $\mathbf{W}_0\!\sim\!\mathcal{U}(-1/d,1/d)$ (uniform sampling) for input dimension $d$ (subsequent sine layers scaled by $1/\omega_0$); hidden sine layers $\mathbf{W}_i\!\sim\!\mathcal{U}(-\sqrt{6/n}/\omega,\sqrt{6/n}/\omega)$ with $n=256$. The SDF output is initialized near zero ($\pm10^{-5}$). ReLU layers in the segmentation head use Kaiming initialization; the final classifier is initialized in $\pm10^{-3}$ so logits start nearly uniform.

\paragraph{Training configuration.}
We use Adam with learning rate $5{\times}10^{-5}$, batch size $1$ (one shape per iteration), and $10$ epochs. Loss weights are
$\lambda_{\text{DM}}{=}7000$, $\lambda_{\text{DNM}}{=}600$, $\lambda_{\text{eik}}{=}50$, $\lambda_{\text{ODW}}{=}10$, $\lambda_{\text{seg}}{=}100$.
The segmentation loss is applied only to labeled manifold samples; geometric losses follow the manifold/non-manifold split.

\paragraph{Segmentation heads.}
Four classifier variants are evaluated with the same SDF trunk:
(i) \texttt{ReLU}: two ReLU layers (with dropout) then linear;
(ii) \texttt{SIREN}: two sine layers (first with large $\omega_0$) then linear;
(iii) \texttt{Hybrid}: one sine layer followed by ReLU layers;
(iv) \texttt{Deep\_skip}: a deeper sine stack with skip connections from the input and an early hidden layer.

\paragraph{Mesh extraction.}
Meshes are extracted with Marching Cubes on a fixed $256^3$ grid over the normalized training volume at the zero level set. SDF queries are chunked to control memory. Vertices (in voxel-index space) are mapped to network space and then to world coordinates via the same normalization; per-vertex labels are obtained by evaluating the segmentation head at \emph{normalized} vertex positions, and each face takes the majority label of its three vertices. Visualizations may use arbitrary color palettes; evaluation uses label metrics (mean Intersection over Union, accuracy, segmentation consistency).

\section{Experiments and Results}

We evaluate on randomly selected meshes from the ABC dataset, with per-face labels generated automatically using PartField. Over-segmentation is intentionally allowed on some shapes to test robustness to varying part counts. PartField is used only for supervision; inference uses our network without PartField components. The test set spans 2--18 parts.

For each test mesh, we load the ground-truth point cloud and labels from the provided \texttt{.ply} file, and sample the same number of points on the predicted mesh surface. All metrics are computed per shape and reported as mean $\pm$ standard deviation across the set.

Reconstruction is evaluated following FlatCAD, using Chamfer distance (CD$_{L1}$ and CD$_{L2}$) computed over bidirectional nearest neighbours, normal consistency (NC) as the mean cosine similarity of matched normals, and micro-averaged F1$_\mu$ at the FlatCAD threshold. Segmentation accuracy (Acc.) is measured by transferring ground-truth labels to predicted samples via nearest neighbour in 3D, computing per-part IoU, aggregating to mean Intersection over Union (mIoU), and reporting per-point accuracy.

We also introduce a segmentation consistency (Consis.) metric. For predicted samples $P=\{(x_i,\ell_i)\}$, we use $\mathbf{1}[\ell_j=\ell_i]$ the indicator function to compute
\[
c_i = \tfrac{1}{k}\sum_{j\in\mathcal{N}_k(x_i)} \mathbf{1}[\ell_j=\ell_i],
\]
with $k=10$, and average $c_i$ over $M=\min(1000,|P|)$ anchors. This measures local label smoothness in prediction space and is invariant to palette permutations. Finally, we record the number of predicted and ground-truth parts for each shape.

\subsection{Results}

Table ~\ref{tab:recon-compact} shows comparison to FlatCAD. We compare four segmentation heads (\texttt{ReLU}, \texttt{SIREN}, \texttt{Hybrid}, \texttt{Deep\_skip}). On the same 20 shapes, SIREN/Hybrid/Deep-skip show no significant changes in CD, NC, or F1 ($\Delta \approx 0$ , paired t-tests $p>0.05$ ), indicating the head is plug-in and reconstruction-neutral; ReLU trends slightly worse.

Table~\ref{tab:main-results} reports the reconstruction and corresponding segmentation mean $\pm$ std over 50 randomly selected shapes. Lower is better for CD$_{L1}$/CD$_{L2}$; higher is better for F1$_\mu$, NC, mIoU, Acc. and Consis. All methods achieve strong reconstruction accuracy. \texttt{SIREN} attains the lowest Chamfer distances, closely followed by \texttt{Hybrid}. \texttt{Hybrid} reaches the best F1$_\mu$ and NC, with \texttt{SIREN} a close second. Segmentation accuracy is uniformly high: mIoU $\geq 0.951$, per-point accuracy $\geq 0.971$, and segmentation consistency $\geq 0.966$. Among models, \texttt{Hybrid} yields the highest mIoU (0.967), while \texttt{SIREN} achieves the best accuracy (0.978) and the highest consistency (0.973).  
Breakdowns by part count show consistent behavior: differences between \texttt{SIREN} and \texttt{Hybrid} remain small, and both clearly outperform \texttt{ReLU} and \texttt{Deep\_skip}.
{Table~\ref{tab:perpart-50}} presents per-part mIoU (± standard deviation). Columns $P_k$ correspond to part label $k$. Beyond per-metric comparisons, we analyze correlations between reconstruction and segmentation metrics (Table~\ref{tab:seg-recon-corr}). Overall, \texttt{SIREN} achieves the lowest Chamfer distances and highest per-point accuracy/consistency, while \texttt{Hybrid} provides the best mIoU and F1$_\mu$/NC. This demonstrates that the proposed framework yields accurate reconstructions with semantically stable segmentations across a range of part counts.
% : CD$_{L1}$ correlates with higher mIoU/Acc., while Consis. shows weak correlation with CD$_{L1}$, suggesting robustness to geometric errors.
% Missing parts are indicated with ``--''. When only one shape contributes, the standard deviation is $0.0000$.
 
\begin{figure*}[t]
  \centering
  \includegraphics[width=\textwidth]{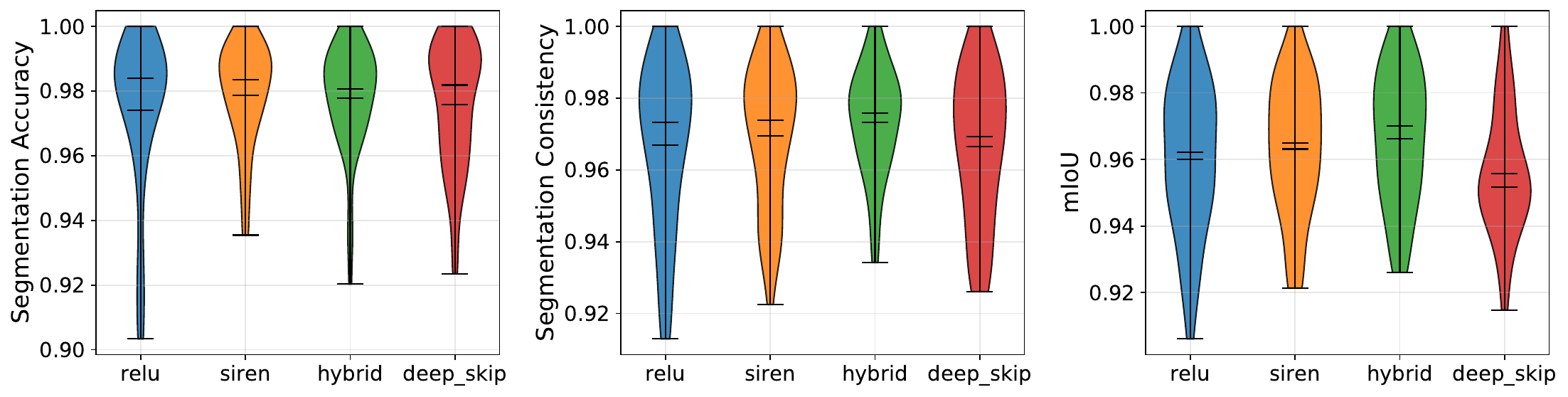}
  \caption{Model comparison across segmentation metrics (colors indicate model types).}
  \label{fig:comparison-colored}
\end{figure*}

\begin{figure*}[t]
  \centering
  \TriptychRow{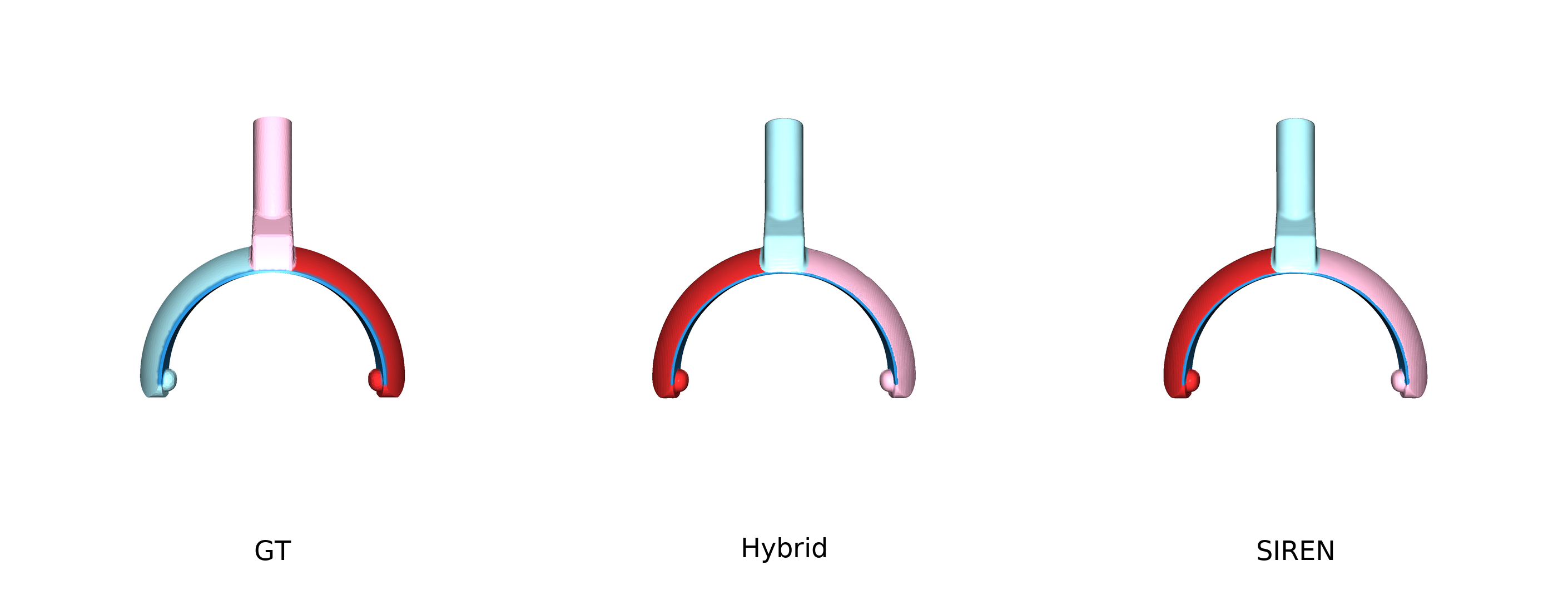}
  \TriptychRow{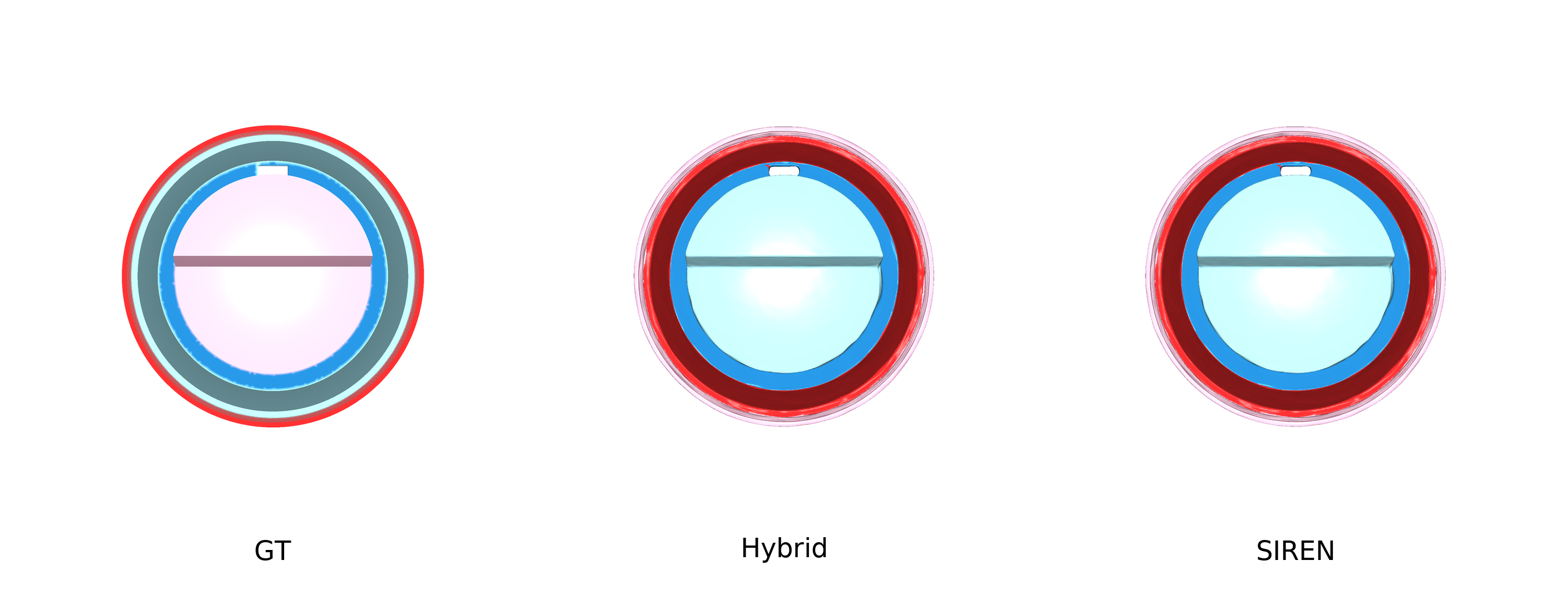}
  % \TriptychRow{036.png}
  \TriptychRow{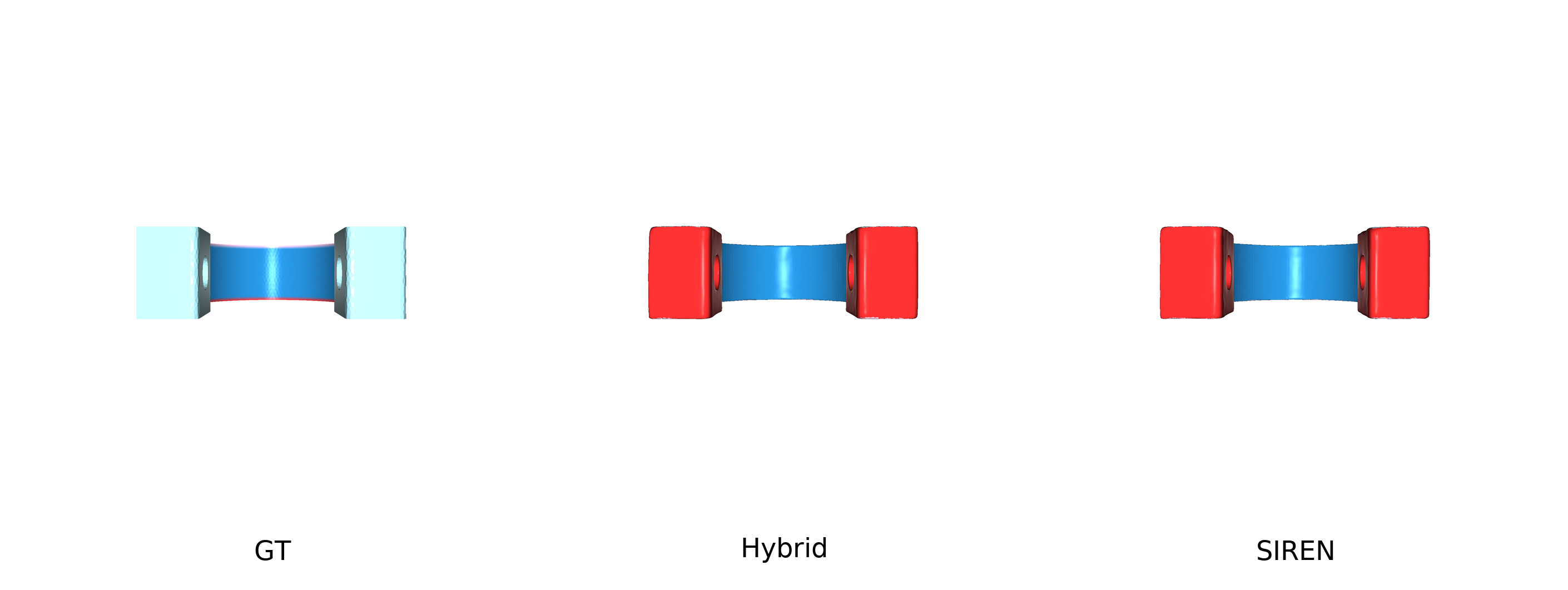}
  \TriptychRow{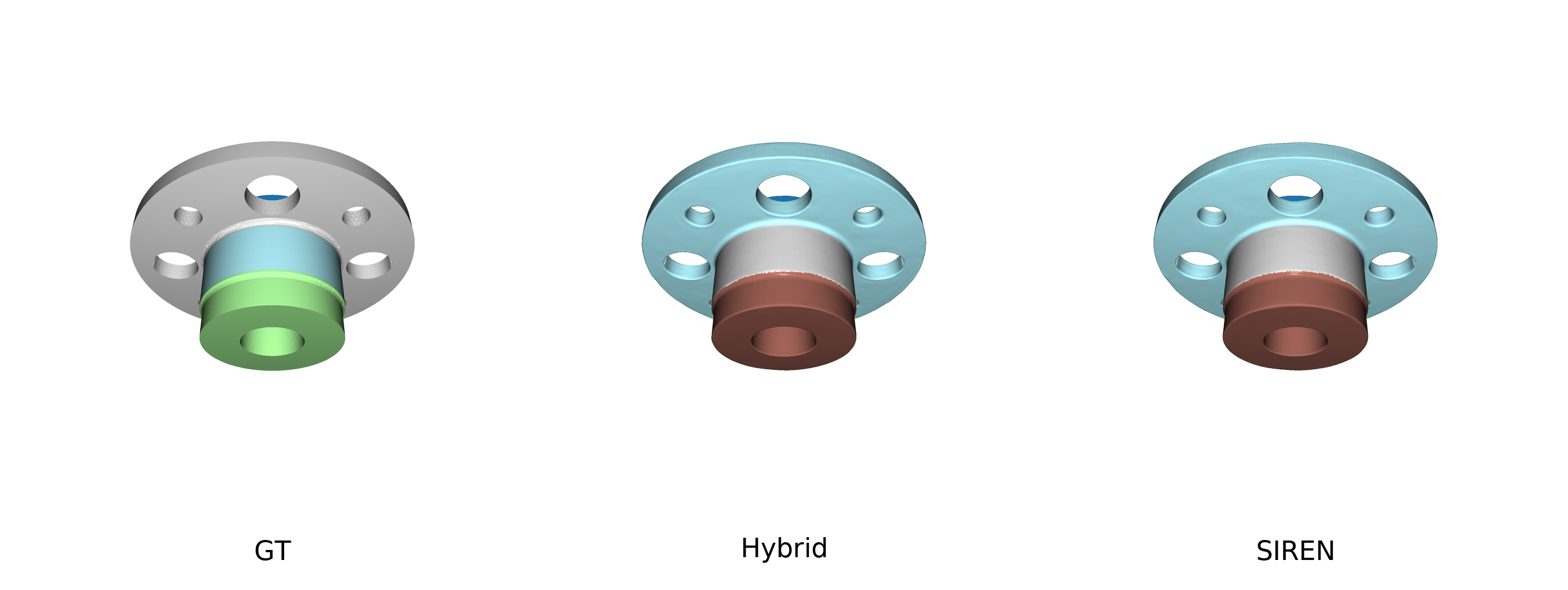}
  \TriptychRow{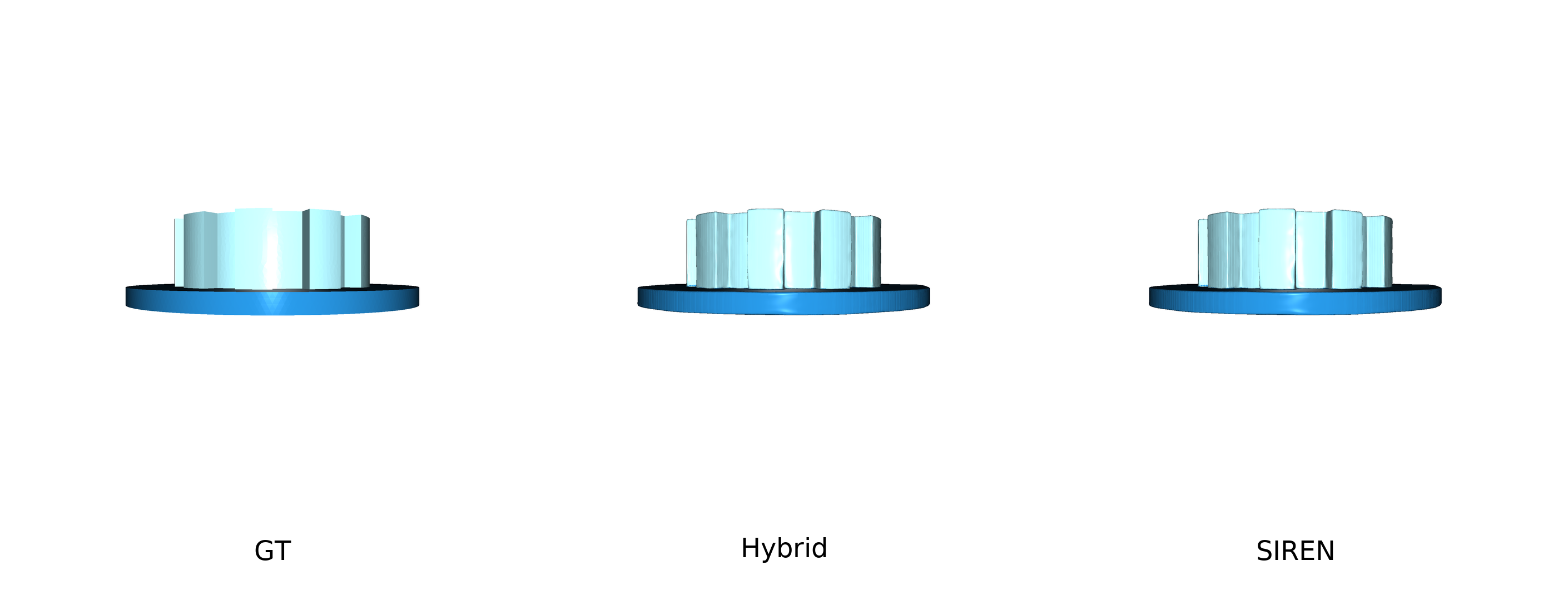}
  \TriptychRow{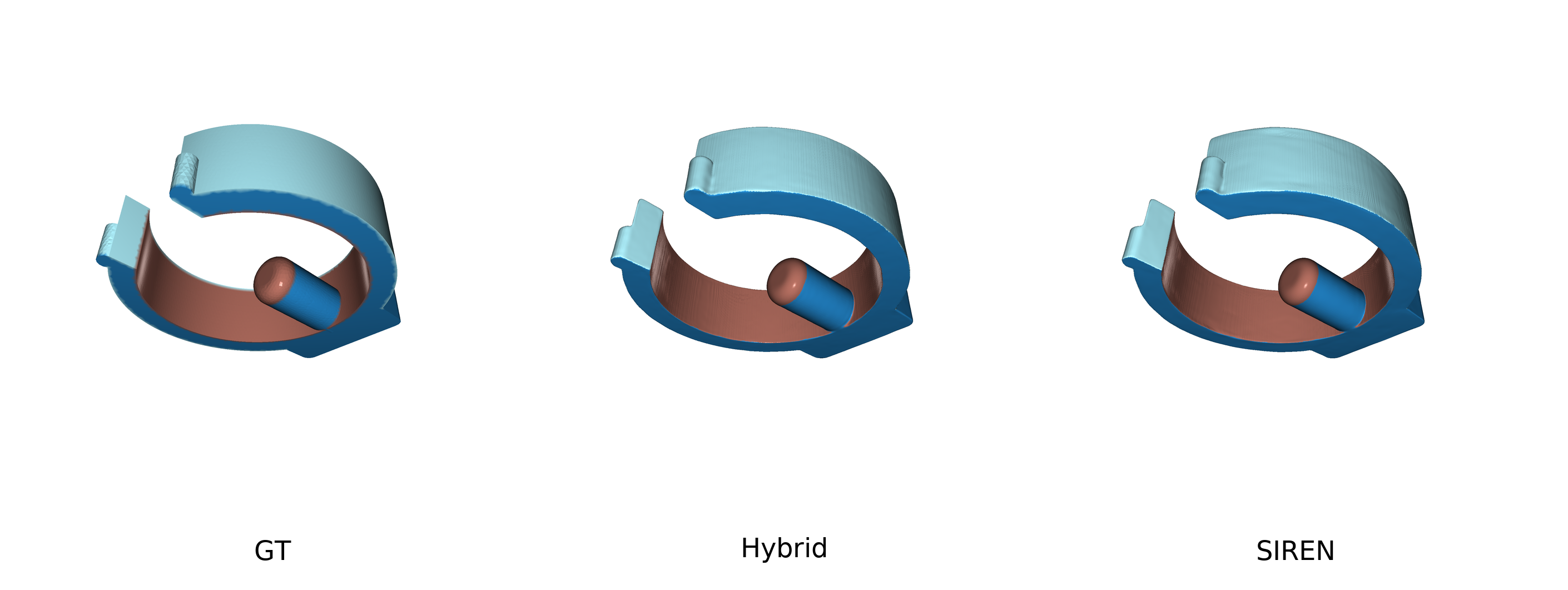}
  % \TriptychRow{triple_153_gt-hybrid-siren.png}
  \caption{GT vs.\ Hybrid vs.\ SIREN qualitative comparison. 
   Colors may differ from the original PartField palette (palette permutations), 
   but \emph{part identities} are evaluated by labels (mIoU/Acc./Consis.), not RGB.}
  \label{fig:triptych-stack}
\end{figure*}

 Figure~\ref{fig:comparison-colored} depicts results across reconstruction and segmentation metrics. Figure~\ref{fig:triptych-stack} shows visual comparisons.

% \noindent\textbf{Takeaway.} \emph{SIREN} achieves the lowest $\mathrm{CD}_{L1}/\mathrm{CD}_{L2}$ and highest Accuracy/Consistency, while \emph{Hybrid} slightly leads in mIoU; the gap between them is small across all part counts (Table~\ref{tab:main-results}).

\subsection{Limitations}

\begin{figure*}[t]
  \centering
  % \TriptychRow{figures/074.png}
  % \TriptychRow{figures/016.png}
  % \TriptychRow{triple_153_gt-hybrid-siren.png}
  \includegraphics[width=0.95\linewidth,trim= 50 60 50 50,clip]{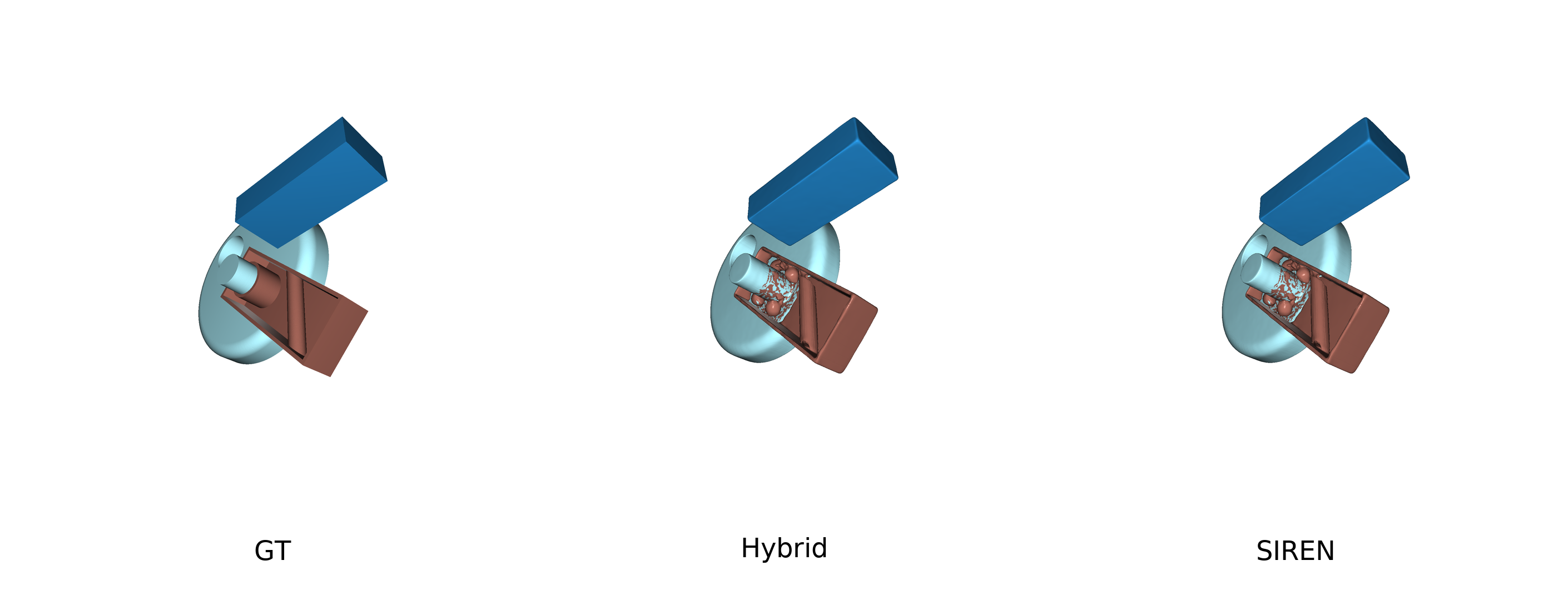}
  \caption{Failure case: GT vs. Hybrid vs. SIREN. The reconstructed meshes are imperfect, 
  yet segmentations remain coherent (correct part count and high consistency).}
  \label{fig:failure}
\end{figure*}

Although the method handles meshes with arbitrary part counts and maintains segmentation stability even with imperfect reconstructions, fine-grained boundary segmentation remains a challenge. We observe occasional leaks or frayed borders in thin structures and at tight fillets (see Fig.~\ref{fig:failure}). This comes from the supervision source: PartField provides per-face labels on meshes of varying tessellation quality, so boundaries inherit face-scale discretization. Higher resolution supervision would mitigate this.

\begin{table}[t]
\centering
\caption{Reconstruction on 20 random shapes. $\Delta$ is head$-$FlatCAD (CD: lower is better; NC/F1$_\mu$: higher).}
\label{tab:recon-compact}
\setlength{\tabcolsep}{4pt}
\begin{tabular}{lcccccc}
\toprule
& \multicolumn{2}{c}{CD$_{L1}$} & \multicolumn{2}{c}{NC} & \multicolumn{2}{c}{F1$_\mu$}\\
\cmidrule(lr){2-3}\cmidrule(lr){4-5}\cmidrule(lr){6-7}
model & mean$\pm$std & $\Delta$ & mean$\pm$std & $\Delta$ & mean$\pm$std & $\Delta$ \\
\midrule
FlatCAD & 0.42$\pm$0.31 & — &
         0.953$\pm$0.042 & — &
         0.849$\pm$0.127 & — \\
ReLU     & 0.57$\pm$0.47 & +0.001489 &
          0.939$\pm$0.062 & -0.0138 &
          0.762$\pm$0.269 & -0.0871 \\
SIREN    & 0.42$\pm$0.28 & +0.000006 &
          0.955$\pm$0.044 & +0.0016 &
          0.834$\pm$0.143 & -0.0150 \\
Hybrid   & 0.42$\pm$0.28 & +0.000009 &
          0.955$\pm$0.044 & +0.0019 &
          0.835$\pm$0.143 & -0.0149 \\
Deep\_skip & 0.45$\pm$0.36 & +0.000361 &
            0.952$\pm$0.048 & -0.0017 &
            0.830$\pm$0.163 & -0.0192 \\
\bottomrule
\end{tabular}
\end{table}

% % ---------------- Reconstruction ----------------
% \begin{table}[h]
%   \centering  
%   \caption{Main results: Reconstruction metrics over test shapes. See Sec. 4 for metric definitions.}
%   \label{tab:main-results-recon}
%   \begin{tabular*}{\linewidth}{@{\extracolsep{\fill}}lccccc}
%     \toprule
%     model\_type & $N$ & CD$_{L1}$ ($\times 10^{-2}$) & CD$_{L2}$ ($\times 10^{-2}$) & NC & F1$_\mu$ \\
%     \midrule
%     ReLU      & 50 & $0.59 \pm 1.30$ & $0.05 \pm 0.21$ & $0.94 \pm 0.07$ & $0.78 \pm 0.28$ \\
%     SIREN     & 50 & $0.51 \pm 0.89$ & $0.04 \pm 0.12$ & $0.95 \pm 0.06$ & $0.82 \pm 0.17$ \\
%     Hybrid    & 50 & $0.49 \pm 0.95$ & $0.03 \pm 0.11$ & $0.95 \pm 0.06$ & $0.83 \pm 0.15$ \\
%     Deep\_skip& 50 & $0.62 \pm 0.65$ & $0.03 \pm 0.13$ & $0.95 \pm 0.06$ & $0.82 \pm 0.19$ \\
%     \bottomrule
%   \end{tabular*}
% \end{table}

% % ---------------- Segmentation ----------------
% \begin{table}[h]
% \ContinuedFloat
%   \centering  
%   \caption{Main results (continued): See Sec. 4 for metric definitions.}
%   \label{tab:main-results-seg}
%   \begin{tabular*}{\linewidth}{@{\extracolsep{\fill}}lcccc}
%     \toprule
%     model\_type & $N$ & mIoU & Acc. & Consis. \\
%     \midrule
%     ReLU       & 50 & $0.96 \pm 0.02$ & $0.97 \pm 0.02$ & $0.97 \pm 0.02$ \\
%     SIREN      & 50 & $0.96 \pm 0.02$ & $0.98 \pm 0.01$ & $0.97 \pm 0.02$ \\
%     Hybrid     & 50 & $0.97 \pm 0.01$ & $0.98 \pm 0.02$ & $0.97 \pm 0.01$ \\
%     Deep\_skip & 50 & $0.95 \pm 0.02$ & $0.98 \pm 0.02$ & $0.97 \pm 0.02$ \\
%     \bottomrule
%   \end{tabular*}
% \end{table}

\begin{table}[t]
  \centering
  \caption{Main results.
  See Sec.~4 for metric definitions.}
  \label{tab:main-results}
  \setlength{\tabcolsep}{4pt}
  \renewcommand{\arraystretch}{1.08}

  %---------------- (a) Reconstruction ----------------
  \begin{minipage}{\linewidth}
    \vspace{2pt}\textbf{(a) Reconstruction}\par\vspace{2pt}
    \begin{tabular*}{\linewidth}{@{\extracolsep{\fill}}lccccc}
      \toprule
      model\_type & $N$ & CD$_{L1}$ ($\times 10^{-2}$) & CD$_{L2}$ ($\times 10^{-2}$) & NC & F1$_\mu$ \\
      \midrule
      ReLU       & 50 & $0.59 \pm 1.30$ & $0.05 \pm 0.21$ & $0.94 \pm 0.07$ & $0.78 \pm 0.28$ \\
      SIREN      & 50 & $0.51 \pm 0.89$ & $0.04 \pm 0.12$ & $0.95 \pm 0.06$ & $0.82 \pm 0.17$ \\
      Hybrid     & 50 & $0.49 \pm 0.95$ & $0.03 \pm 0.11$ & $0.95 \pm 0.06$ & $0.83 \pm 0.15$ \\
      Deep\_skip & 50 & $0.62 \pm 0.65$ & $0.03 \pm 0.13$ & $0.95 \pm 0.06$ & $0.82 \pm 0.19$ \\
      \bottomrule
    \end{tabular*}
  \end{minipage}

  \vspace{0.6em}

  %---------------- (b) Segmentation ----------------
  \begin{minipage}{\linewidth}
    \vspace{2pt}\textbf{(b) Segmentation}\par\vspace{2pt}
    \begin{tabular*}{\linewidth}{@{\extracolsep{\fill}}lcccc}
      \toprule
      model\_type & $N$ & mIoU & Acc. & Consis. \\
      \midrule
      ReLU       & 50 & $0.96 \pm 0.02$ & $0.97 \pm 0.02$ & $0.97 \pm 0.02$ \\
      SIREN      & 50 & $0.96 \pm 0.02$ & $0.98 \pm 0.01$ & $0.97 \pm 0.02$ \\
      Hybrid     & 50 & $0.97 \pm 0.01$ & $0.98 \pm 0.02$ & $0.97 \pm 0.01$ \\
      Deep\_skip & 50 & $0.95 \pm 0.02$ & $0.98 \pm 0.02$ & $0.97 \pm 0.02$ \\
      \bottomrule
    \end{tabular*}
  \end{minipage}
\end{table}

\begin{table*}[t]
  \centering
  \caption{Per-part IoU on 50 shapes (mean$\pm$std). Columns P$_k$ correspond to part label $k$.}
  \label{tab:perpart-50}
  \setlength{\tabcolsep}{3.5pt}
  \subcaptionbox{P$_0$–P$_8$ \label{tab:perpart-50a}}{%
    \resizebox{\linewidth}{!}{
\begin{tabular}{lccccccccc}
\toprule
model\_type & P0 & P1 & P2 & P3 & P4 & P5 & P6 & P7 & P8 \\
\midrule
ReLU & 0.93 $\pm$ 0.20 & 0.94 $\pm$ 0.12 & 0.94 $\pm$ 0.12 & 0.93 $\pm$ 0.17 & 0.90 $\pm$ 0.23 & 0.89 $\pm$ 0.24 & 0.88 $\pm$ 0.22 & 0.89 $\pm$ 0.15 & 0.92 $\pm$ 0.09 \\
SIREN & 0.94 $\pm$ 0.17 & 0.94 $\pm$ 0.15 & 0.94 $\pm$ 0.16 & 0.93 $\pm$ 0.13 & 0.95 $\pm$ 0.14 & 0.90 $\pm$ 0.28 & 0.91 $\pm$ 0.20 & 0.92 $\pm$ 0.15 & 0.98 $\pm$ 0.01 \\
Hybrid & 0.94 $\pm$ 0.14 & 0.93 $\pm$ 0.17 & 0.96 $\pm$ 0.12 & 0.97 $\pm$ 0.08 & 0.97 $\pm$ 0.10 & 0.95 $\pm$ 0.13 & 0.94 $\pm$ 0.18 & 0.92 $\pm$ 0.16 & 0.99 $\pm$ 0.02 \\
Deep\_skip & 0.94 $\pm$ 0.14 & 0.94 $\pm$ 0.14 & 0.95 $\pm$ 0.11 & 0.94 $\pm$ 0.12 & 0.95 $\pm$ 0.14 & 0.89 $\pm$ 0.24 & 0.90 $\pm$ 0.16 & 0.91 $\pm$ 0.18 & 0.98 $\pm$ 0.01 \\
\bottomrule
\end{tabular}}}
  \vspace{2pt}

  \subcaptionbox{P$_9$–P$_{17}$ \label{tab:perpart-50b}}{%
    \resizebox{\linewidth}{!}{\begin{tabular}{lccccccccc}
\toprule
model\_type & P9 & P10 & P11 & P12 & P13 & P14 & P15 & P16 & P17 \\
\midrule
ReLU & 0.89 $\pm$ 0.03 & 0.92 $\pm$ 0.07 & 0.88 $\pm$ 0.17 & 0.99 $\pm$ 0.00 & 0.72 $\pm$ 0.15 & 0.99 $\pm$ 0.00 & 0.99 $\pm$ 0.00 & 0.98 $\pm$ 0.00 & 0.99 $\pm$ 0.00 \\
SIREN & 0.91 $\pm$ 0.09 & 0.94 $\pm$ 0.08 & 0.92 $\pm$ 0.14 & 0.94 $\pm$ 0.06 & 0.75 $\pm$ 0.14 & 0.98 $\pm$ 0.00 & 0.99 $\pm$ 0.00 & 0.99 $\pm$ 0.00 & 0.98 $\pm$ 0.00 \\
Hybrid & 0.95 $\pm$ 0.07 & 0.99 $\pm$ 0.01 & 0.96 $\pm$ 0.09 & 0.99 $\pm$ 0.01 & 0.90 $\pm$ 0.07 & 0.99 $\pm$ 0.00 & 0.99 $\pm$ 0.00 & 0.99 $\pm$ 0.00 & 0.99 $\pm$ 0.00 \\
Deep\_skip & 0.85 $\pm$ 0.11 & 0.93 $\pm$ 0.02 & 0.91 $\pm$ 0.17 & 0.96 $\pm$ 0.04 & 0.81 $\pm$ 0.07 & 0.98 $\pm$ 0.00 & 0.98 $\pm$ 0.00 & 0.98 $\pm$ 0.00 & 0.98 $\pm$ 0.00 \\
\bottomrule
\end{tabular}}}
\end{table*}

% \begin{table}[t]
% \centering
% \caption{Reconstruction comparison (mean$\pm$std) across FlatCAD (`flat`) and our four heads on the same 13 shapes. Lower is better for CD; higher is better for NC/F1$_\mu$.}
% \label{tab:flat-vs-heads-mean}
% \begin{tabular}{lcccc}
% \toprule
% model & CD$_{L1}$ & CD$_{L2}$ & NC & F1$_\mu$ \\
% \midrule
% flat & 0.004216 $\pm$ 0.003115 & 0.000156 $\pm$ 0.000466 & 0.953191 $\pm$ 0.042106 & 0.849442 $\pm$ 0.127430 \\
% relu & 0.005705 $\pm$ 0.004768 & 0.000198 $\pm$ 0.000454 & 0.939393 $\pm$ 0.062042 & 0.762344 $\pm$ 0.268947 \\
% siren & 0.004223 $\pm$ 0.002821 & 0.000155 $\pm$ 0.000464 & 0.954821 $\pm$ 0.044002 & 0.834439 $\pm$ 0.142920 \\
% hybrid & 0.004225 $\pm$ 0.002826 & 0.000155 $\pm$ 0.000465 & 0.955126 $\pm$ 0.043715 & 0.834508 $\pm$ 0.143170 \\
% deep\_skip & 0.004577 $\pm$ 0.003619 & 0.000174 $\pm$ 0.000507 & 0.951526 $\pm$ 0.048395 & 0.830287 $\pm$ 0.162869 \\
% \bottomrule
% \end{tabular}
% \end{table}

% \begin{table}[h]
%   \centering
%   \tablesetup
%   \caption{Per-part IoU on 50 shapes (mean$\pm$std). Columns P$_k$ correspond to part label $k$.}
%   \label{tab:part-iou-a}
%   \begin{adjustbox}{width=\linewidth}
%     \input{5_supp_part_iou_p0p8}
%   \end{adjustbox}
% \end{table}

% \begin{table}[h]
%   \ContinuedFloat
%   \centering
%   \tablesetup
%   \caption{(continued) Per-part IoU (P$_6$–P$_{11}$).}
%   \begin{adjustbox}{width=\linewidth}
%     \input{5_supp_part_iou_p9p17}
%   \end{adjustbox}
% \end{table}

%--- Correlation between reconstruction and segmentation ---

\begin{table}[t]
  \centering
  \tablesetup
  \caption{Correlation (Pearson $r$) between reconstruction metrics and segmentation metrics. 
  Negative values for CD$_{L1}$ indicate that lower Chamfer distances correlate with higher segmentation scores; 
  correlations with Consis.\ are weak, suggesting robustness to geometric errors.}
  \label{tab:seg-recon-corr}
  \begin{adjustbox}{width=\linewidth}
    \begin{tabular*}{\linewidth}{@{\extracolsep{\fill}}lccccc}
\toprule
 & Overall & ReLU & SIREN & Hybrid & Deep\_skip \\\midrule
CD$_{L1}$ $\leftrightarrow$ mIoU & -0.68 & -0.74 & -0.77 & -0.68 & -0.77 \\
CD$_{L1}$ $\leftrightarrow$ Acc. & -0.76 & -0.78 & -0.82 & -0.67 & -0.85 \\
CD$_{L1}$ $\leftrightarrow$ Consis. & 0.08 & -0.12 & -0.13 & -0.06 & -0.19 \\
NC $\leftrightarrow$ mIoU & 0.23 & 0.27 & 0.27 & 0.19 & 0.20 \\
NC $\leftrightarrow$ Acc. & 0.25 & 0.28 & 0.26 & 0.22 & 0.19 \\
NC $\leftrightarrow$ Consis. & 0.13 & 0.10 & 0.24 & 0.03 & 0.16 \\
F1$_{\mu}$ $\leftrightarrow$ mIoU & 0.48 & 0.65 & 0.38 & 0.41 & 0.40 \\
F1$_{\mu}$ $\leftrightarrow$ Acc. & 0.57 & 0.73 & 0.54 & 0.59 & 0.56 \\
F1$_{\mu}$ $\leftrightarrow$ Consis. & 0.10 & 0.18 & 0.11 & 0.09 & 0.07 \\
\bottomrule
\end{tabular*}

  \end{adjustbox}
\end{table}

% --- Segmentation versus reconstruction strata ---

% \begin{table}[t]
%   \centering
%   \tablesetup
%   \caption{Segmentation quality in best vs.\ worst reconstruction quartiles (per model, stratified by CD$_{L1}$). 
%   Best = lowest 25\% CD$_{L1}$; Worst = highest 25\%. Differences are small, especially for Consis., 
%   indicating that the proposed consistency measure is relatively insensitive to mild surface errors. $N_{\text{best}}$ and $N_{\text{worst}}$ denote the number of shapes per model falling into the best/worst quartiles (here 4 each out of 16).}
%   \label{tab:seg-vs-recon-strata}
%   \begin{adjustbox}{width=\linewidth}
%     \input{5_seg_vs_recon_strata.tex}
%   \end{adjustbox}
% \end{table}

\section{Conclusions}
We presented a joint implicit framework for CAD surface reconstruction and part segmentation by extending FlatCAD with a segmentation head. The approach leverages shared SIREN features to predict both signed distance values and per-part labels in a unified representation. Experiments on PartField-supervised ABC meshes demonstrate that the method achieves strong reconstruction accuracy while delivering high segmentation performance, with mIoU above 0.96 and accuracy above 0.97 across diverse part counts. A paired comparison with FlatCAD shows \emph{no significant differences} in reconstruction, indicating that the head is plug-in and reconstruction-neutral. Our proposed segmentation consistency metric further confirms label coherence even under imperfect reconstructions. Because the head is decoupled from the SDF trunk, it can be attached to \emph{existing} SDF-based reconstruction models with minimal effort.

% Limitations remain in boundary precision due to mesh-level supervision, suggesting future work on boundary-aware training and higher-resolution annotations.
% perform the experiments and collect the data (primiary priority)
% \input{4_experiments}

% \FloatBarrier --> dont use it
% \bibliography{related_work_bib}
\bibliographystyle{splncs04}
\bibliography{main}
% \nocite{*}
\end{document}